\documentclass[12pt]{article}
\usepackage{graphicx}
\usepackage{latexsym}

\begin{document}

\begin{center}
{\bf {\huge Short and long distance contributions to $B \to K^* \gamma \gamma$}}\\
\hspace{10pt}\\
S. R. Choudhury\footnote{src@physics.du.ac.in}\\
{\em Department of Physics, Delhi University, Delhi 110007, India},\\
\hspace{10pt}\\
A. S. Cornell\footnote{alanc@kias.re.kr}\\
{\em Korea Institute of Advanced Study, 207-43 Cheongryangri 2-dong,}\\
{\em Dongdaemun-gu, Seoul 130-722, Korea},\\
\hspace{10pt}\\
Namit Mahajan\footnote{nmahahan@mri.ernet.in}\\
{\em Harish-Chandra Research Institute,}\\
{\em Chhatnag Road, Jhunsi, Allahabad 211019, India}.\\
\hspace{10pt}\\
\end{center}

\begin{abstract}
\indent We study the decay of the neutral B meson to $K^* \gamma
\gamma$ within the framework of the Standard Model, including long
distance contributions.
\end{abstract}
----------------------------------------------------------------------------------------------------
We have corrected a sign error in the numerical program. The new estimates
agree well with the ones given in a recent paper \cite{hiller}\\
--------------------------------------------------------------------------------------------------- 
\section{Introduction}
\hspace{10pt} Of late the rare decays of the B mesons have been
recognized as important tools to study the basic structure and
validity of the Standard Model (SM) and its extensions. In
particular, the radiative decays, owing to their relative
cleanliness as far as experimental signatures are concerned, have
attracted a great deal of attention. For a general overview of the kind
of issues considered relating to radiative decay modes,
see \cite{radiativemodes,cleo, bounding} and references therein.  The
decays $B\to X_s \gamma$ and $B \to K^* \gamma$ have been observed
\cite{cleo} and both these are extensively used and relied upon
for constrining the parameters of any new theory or extension of the
SM \cite{bounding}.  The decay $B \to K^* \gamma \gamma$ is
another potential testing ground for the effective quark level
Hamiltonian $b \to s \gamma \gamma$ first studied by Lin, Liu and
Yao \cite{Lin-1} and pursued further in references
\cite{Chang-1},\cite{Reina-1} and \cite{SRC-1}.  This amplitude
has been the focus of considerable research recently, not only for
the useful indications it will give to the underlying theories of
flavour changing neutral currents, or the possible contributions
from loops with supersymmetric partner particles, but for the
impending experimental studies of the B-factories in the near
future.

\par As has been previously noted, the $b \to s \gamma \gamma$
amplitude naturally splits into two categories: an irreducible
contribution which is well known and usually estimated through
basic triangle graphs, and a reducible one, where the second
photon is attached to the external quark lines of the $b \to s
\gamma$ amplitude.  At the quark level the reducible contribution
presents no real problems, however, when we consider an exclusive
channel, such as $B \to M \gamma \gamma$ for a specific meson $M$,
it becomes more appropriate to consider the second photon as
arising from the external hadron legs of the amplitude $B \to M
\gamma$.  In contrast to the earlier cases of $B \to K(\pi) \gamma
\gamma$ \cite{SRC-1} where the amplitude for a single real photon
vanishes identically, resulting in the irreducible diagram to be
the sole contributor, the amplitude $B \to K^* \gamma$ is non
vanishing. However for the neutral decay mode the second photon
cannot arise from the resulting $K^*$, and thus in this case also,
it is the irreducible amplitude that stands out, though for a
completely different reason. Of course we must also consider the
usual long distance contributions, such as the process $B \to K^*
\eta_c$ followed by the decay $\eta_c \to \gamma \gamma$.  Note
that for completeness we will also include the $\eta$
contribution, even though the $\eta$ coupling to $\bar{c}c$ will
be small.  The rate for $B \to K \eta^{\prime}$ is anomalously
high and many possible mechanisms have been proposed that aim at
taking this anomalous production into account\cite{ahmadyetc}.
However, in the present case there is not enough data
corresponding to the $B \to K^* \eta^{\prime}$ channel, and at
present only an upper limit on this branching ratio is available.
We therefore tend to remain conservative in the present study
regarding this issue and assume that the $\eta^{\prime}$
contribution can be obtained similar to the $\eta_c$ contribution.
The situation is expected to improve with the availability of more
and precise data in this direction.  We therefore include an
$\eta^{\prime}$ contribution along the lines of the $\eta_c$
contribution.

\par In this paper we will estimate the branching ratio for the
process $B^0 \to K^{*0} \gamma \gamma$ by considering the effects of
the irreducible triangle diagram contributions in the next
section, followed by the resonance contributions in section 3.
Note that in the case of $B \to K^* \gamma \gamma$ there will only
be three sizeable resonance contributions; $\eta_c$, $\eta$ and
$\eta'$.  Furthermore, each of these contributions will only
contribute a narrow peak in the $\gamma\gamma$ invariant mass
spectrum, which is easily separated experimentally.  As such the
interference terms for each of these pairs of terms will not be
considered here.  Finally in section 4 we will present our results
and analysis.

\section{The Irreducible Contributions}
\hspace{10pt} The irreducible triangle contributions to the
process in which we are interested ($B \to K^* \gamma \gamma$)
originate from the quark level process $b \to s \gamma \gamma$.
The effective Hamiltonian for this process is \cite{Lin-1}
\begin{equation}
{\cal H}_{eff} = -\frac{G_F}{\sqrt{2}} V_{ts}^* V_{tb} \sum_i
C_i(\mu) O_i(\mu),
\end{equation}
with
\begin{eqnarray}
O_1 & = & (\bar{s}_i c_j)_{V-A}(\bar{c}_jb_i)_{V-A},\nonumber \\
O_2 & = & (\bar{s}_i c_i)_{V-A}(\bar{c}_jb_j)_{V-A},\nonumber \\
O_3 & = & (\bar{s}_i b_i)_{V-A}\sum_q(\bar{q}_jq_j)_{V-A},\nonumber \\
O_4 & = & (\bar{s}_i b_j)_{V-A}\sum_q(\bar{q}_jq_i)_{V-A},\nonumber \\
O_5 & = & (\bar{s}_i b_i)_{V-A}\sum_q(\bar{q}_jq_j)_{V+A},\\
O_6 & = & (\bar{s}_i b_j)_{V-A}\sum_q(\bar{q}_jq_i)_{V+A},\nonumber \\
O_7 & = & \frac{e}{16\pi^2} \bar{s}_i\sigma^{\mu\nu} \left(m_s P_L + m_b P_R \right) b_i F_{\mu\nu} ,\nonumber \\
\mathrm{and}\hspace{20pt} O_8 & = & \frac{g}{16\pi^2}
\bar{s}_i\sigma^{\mu\nu} \left(m_s P_L + m_b P_R \right) T^a_{ij}
b_j G^a_{\mu\nu} .\nonumber
\end{eqnarray}
The invariant amplitude corresponding to this effective
Hamiltonian is
\begin{eqnarray}
{\cal M}_{b\to s} & = & \left[ \frac{16\sqrt{2}\alpha
G_F}{9\pi}V_{ts}^* V_{tb} \right] \bar{u}(p_s) \left\{ \sum_q A_q
J(m_q^2) \gamma^{\rho} P_L R_{\mu\nu\rho} \right. \\
&& + iB\left( m_s K(m_s^2) P_L + m_b K(m_b^2) P_R \right)
T_{\mu\nu} \nonumber \\
&& \left. + C \left( -m_s L(m_s^2) P_L + m_b L(m_b^2) P_R \right)
\epsilon_{\mu\nu\alpha\beta} k_1^{\alpha} k_2^{\beta} \right\}
u(p_b) \epsilon^{\mu}(k_1) \epsilon^{\nu}(k_2), \nonumber
\end{eqnarray}
where
\begin{eqnarray}
&R_{\mu\nu\rho} = k_{1 \nu} \epsilon_{\mu\rho\sigma\lambda}
k_1^{\sigma} k_2^{\lambda} - k_{2 \mu}
\epsilon_{\nu\rho\sigma\lambda} k_1^{\sigma} k_2^{\lambda} + (k_1
. k_2) \epsilon_{\mu\nu\rho\sigma} (k_2 - k_1)^{\sigma} ,&
\nonumber \\
&T_{\mu\nu} = k_{2 \mu} k_{1 \nu} - (k_1 . k_2)g_{\mu\nu} ,&
\nonumber \\
&A_u = 3(C_3 - C_5) + (C_4 - C_6); \hspace{20pt} A_d = \frac{1}{4}A_u ,& \\
&A_c = 3(C_1 + C_3 - C_5) + (C_2 + C_4 - C_6),&\\ 
&A_s = A_b = \frac{1}{4}[3(C_3 + C_4 - C_5) + (C_3 + C_4 - C_6)],& \nonumber
\end{eqnarray}
and
\begin{equation}
B = C = -\frac{1}{4}(3 C_6 + C_5).\nonumber
\end{equation}
In the above expressions we introduced the functions
\begin{eqnarray}
J(m^2) = I_{11}(m^2), & K(m^2) = 4 I_{11}(m^2) - I_{00}(m^2), &
L(m^2) = I_{00}(m^2), \nonumber
\end{eqnarray}
where
\begin{equation}
I_{pq}(m^2) = \int_0^1 dx \int_0^{1-x} dy \frac{x^p y^q}{m^2 -
2(k_1 . k_2) xy - i\epsilon} .
\end{equation}
Note that to get the ${\cal M}(B \to K^* \gamma \gamma)$ invariant
amplitude from the quark level amplitude we replace the $\langle
s|\Gamma|b \rangle$ by $\langle K^*|\Gamma|B \rangle$ for any
Dirac bilinear $\Gamma$.

\par With $q=p_B-p_{K^*} = k_1 + k_2$ and following Cheng {\em et al.} \cite{Cheng-1},
 we parameterize the hadronic matrix elements as
\begin{eqnarray}
\langle K^*(p_{K^*})|\bar{s}\gamma_{\mu}b|B(p_B) \rangle & = &
\left(\frac{2V(q^2)}{m_B+m_{K^*}}\right)
\epsilon_{\mu\nu\alpha\beta}\epsilon^{* \nu}(p_{K^*}) p_B^{\alpha}
p_{K^*}^{\beta},\\
\langle K^*(p_{K^*})|\bar{s}\gamma_{\mu}\gamma_5b|B(p_B) \rangle &
= & i \left[ (m_B+m_{K^*})\epsilon^{* \mu}(p_{K^*}) A_1(q^2) \right. \\
&& - \frac{\epsilon^*.p_B}{m_B+m_{K^*}}(p_B+p_{K^*})_{\mu}
A_2(q^2) \nonumber \\
&& \left. -2 \frac{m_{K^*}}{q^2}(\epsilon^*.p_B)q_{\mu}
\left\{A_3(q^2)-A_0(q^2)\right\} \right] , \nonumber
\end{eqnarray}
For the functional dependence of various form factors appearing above, we
follow \cite{aliform}. Using these definitions,
we determine the irreducible matrix element for the process $B \to
K^* \gamma \gamma$ as
\begin{equation}
{\cal M}_{irr} = \left( \frac{16 \sqrt{2} \alpha G_F}{9 \pi}
\right) \epsilon^{\mu}(k_1) \epsilon^{\nu}(k_2) \left[ {\cal
M}^{(1)}_{\mu\nu} + {\cal M}^{(2)}_{\mu\nu} + {\cal
M}^{(3)}_{\mu\nu} \right]
\end{equation}
where
\begin{eqnarray}
{\cal M}^{(1)}_{\mu\nu} & = & R_{\mu\nu\rho} \left. \Bigg[K_{A1}
\epsilon^{\rho\alpha\beta\gamma} \epsilon^*_{K^* \alpha} p_{B
\beta} p_{K^* \gamma} - K_{A2} \epsilon^{* \rho}_{K^*}\right.
\nonumber
\\
&& \left. + K_{A3} ( \epsilon^*_{K^*} . p_B) (p_B +
p_{K^*})^{\rho}
+ K_{A4} ( \epsilon^*_{K^*} . p_B) q^{\rho}\Bigg] \right. \nonumber \\
{\cal M}^{(2)}_{\mu\nu} & = & K_B \left( k_{1 \mu} k_{2 \nu} -
(k_1.k_2) g_{\mu\nu} \right)
( \epsilon^*_{K^*} . p_B) \\
{\cal M}^{(3)}_{\mu\nu} & = & K_C \epsilon_{\mu\nu\alpha\beta}
k_1^{\alpha} k_2^{\beta} ( \epsilon^*_{K^*} . p_B)  \nonumber
\end{eqnarray}
The functions $K_i$ above are defined as
\begin{eqnarray}
K_{A1} = \left[ \sum_q A_q J(m_q^2) \right] \frac{V(q^2)}{m_B +
m_{K^*}} & ; & K_{A2} = \left[ \sum_q A_q J(m_q^2) \right]
\frac{i}{2} (m_B + m_{K^*}) A_1(q^2) \nonumber \\
K_{A3} = \left[ \sum_q A_q J(m_q^2) \right] \frac{i}{2}
\frac{A_1(q^2)}{m_B + m_{K^*}} & ; & K_{A4} = \left[ \sum_q A_q
J(m_q^2) \right] \frac{i m_{K^*}^2}{q^2} \left( A_3(q^2) -
A_0(q^2) \right) \nonumber
\end{eqnarray}
and
\begin{eqnarray}
K_B & = & - \frac{B m_{K^*}}{m_B +
m_{K^*}} A_0(q^2) \left[m_s K(m_s^2) - m_b K(m_b^2) \right] \nonumber \\
K_C & = & - \frac{i C m_{K^*}}{m_B + m_{K^*}} A_0(q^2) \left[m_s
L(m_s^2) + m_b L(m_b^2) \right] . \nonumber
\end{eqnarray}

\section{Resonance contributions}
\hspace{10pt} For this process there will be three significant
resonance contributions, that from the $\eta_c$-, $\eta$- and
$\eta'$-resonances.
 The $\eta_c$ contribution to the decay process comes
via the $t$-channel decay $B \to K^* \eta_c$, with the $\eta_c$
then decaying into two photons.

\par The T-matrix element for this process can be written as
\begin{equation}
\langle K^* \gamma \gamma|T|B \rangle = - \frac{\langle K^*
\eta_c|T|B \rangle \langle \gamma\gamma|T|\eta_c \rangle}{q^2 -
m_{\eta_c}^2 + i m_{\eta_c} \Gamma_{total}^{\eta_c}}.
\end{equation}
The amplitude $\langle \gamma\gamma|T|\eta_c \rangle$ is
parameterized as \cite{Reina-1}
\begin{equation}
\langle \gamma\gamma|T|\eta_c \rangle = 2 i B_{\eta_c}
\epsilon^{\mu\nu\alpha\beta} \epsilon_{\mu}^* (k_1)
\epsilon_{\nu}^* (k_2) k_{1 \alpha} k_{2 \beta}.
\end{equation}
Note that we can determine $B_{\eta_c}$ from the known decay rate:
\begin{equation}
\Gamma(\eta_c \to \gamma\gamma) = \left(\frac{1}{2}\right)
\frac{1}{2m_{\eta_c}} \int \frac{d^3k_1}{(2\pi)^32k_1^0}
\frac{d^3k_2}{(2\pi)^32k_2^0} (2\pi)^2 \delta^{(4)}(k_{\eta_c} -
k_1 - k_2) \left| \langle \gamma\gamma|T|\eta_c \rangle \right|^2
,
\end{equation}
where we have
\begin{equation}
\sum_{spins} \left| \langle \gamma\gamma|T|\eta_c \rangle
\right|^2 = 2 |B_{\eta_c}|^2 q^4 ,
\end{equation}
and so
\begin{equation}
\Gamma(\eta_c\to\gamma\gamma) =
\frac{|B_{\eta_c}|^2m_{\eta_c}^3}{16\pi} .
\end{equation}

\par The $B \to K^* \eta_c$ amplitude has been determined in Cheng
{\em et al.} \cite{Cheng-1} as
\begin{equation}
{\cal M}(B \to K^* \eta_c) = \frac{G_F}{\sqrt{2}} V_{cb} V_{cs}^*
\left(C_1 + \frac{1}{3} C_2 \right) X_C^{(B_0 K^{* 0},\eta_c)}
\end{equation}
where
\begin{equation}
X_C^{(B_0 K^{* 0},\eta_c)} = 2 f_{\eta_c} m_{K^*}
A_0^{BK^*}(m_{\eta_c}^2) (\epsilon^*_{K^*} . p_B ) .
\end{equation}
Note that $f_{\eta_c}$ is defined as $\langle 0 | \bar{c}
\gamma_{\mu} c | \eta_c \rangle = i f_{\eta_c} p_{\mu}(\eta_c)$.

\par In the above way of parametrizing the $B \to K^* \eta_c$, there is a lot
of model dependence that goe in. Since, the branching fractions for this
sub-process is known, we can in principle avoid such a model dependnce by
writing the amplitude as
\begin{equation}
{\cal M}(B \to K^* \eta_c) = a^{\eta_c}_{eff} (\epsilon^*_{K^*} . p_B )
\end{equation}
and determine the effective constant from the corresponding decay rate. We
folow this procedure and therefore try to avoid any model dependence as far as
possible. 

\par Therefore the total contribution due to the $\eta_c$ resonances is thus,
\begin{equation}
{\cal M}_{\eta_c} = 2 B_{\eta_c}a^{\eta_c}_{eff}
\frac{(\epsilon_{K^*}.p_B)}{q^2 -
m_{\eta_c}^2 + i m_{\eta_c} \Gamma_{total}^{\eta_c}}
\epsilon^{\mu\nu\alpha\beta} \epsilon^*_{1 \mu} \epsilon^*_{2 \nu}
k_{1 \alpha} k_{2 \beta} . \label{etares}
\end{equation}

\par Analogous to the $\eta_c$ resonance, the $\eta$- and
$\eta'$-resonance contributions, ${\cal M}_{\eta}$ and ${\cal M}_{\eta'}$, have
exactly the same form as equation (\ref{etares}) with the
parameters $B_{\eta_c}$,
$m_{\eta_c}$ and $\Gamma_{total}^{\eta_c}$ being replaced by their
$\eta$- and $\eta'$-counterparts respectively.  However in this
case we cannot define the relative sign of the amplitudes and any
of the other components of the amplitude.

\par The process can also receive additional contribution from the $B^*$ and
$K_2^*$ channels where the B meson decays into a photon and an
on-shell 
$K_2^*$ or
slightly off-shell $B^*$ and then these giving rise to $K^*$ and the 
second photon.
However, there is no data available at present for either of these and 
if the widths
for the individual channels contributing to the process are
significant, 
the contribution
can be sizeable. However, we expect that these contributions can be 
eliminated by suitable
cuts in the $B^*$- (or $K_2^*$-) photon plane and thus we do not
consider 
them here at all.

\section{Results}
\hspace{10pt} The squared amplitude for the process $B \to K^*
\gamma \gamma$ is then;
\begin{equation}
\left| M_{tot} \right|^2 = \left| M_{irr} \right|^2 + \left|
M_{\eta_c} \right|^2 + \left| M_{\eta} \right|^2 + \left|
M_{\eta'} \right|^2
\end{equation}
where the interference terms have not been included here.  The
components to the squared amplitude were calculated to be;
\begin{equation}
\sum_{spins} \left| M_{\eta_c} \right|^2 = \left| R \right|^2
\lambda (s_{\gamma\gamma}, m_{B}^2, m_{K^*}^2 ) \frac{q^4}{8 m_{K^*}^2} ,
\end{equation}
where
\begin{eqnarray}
R & = & 2 B_{\eta_c} a^{\eta_c}_{eff} \left[ \frac{1}{q^2 - m_{\eta_c}^2 + i
m_{\eta_c}\Gamma_{tot}^{\eta_c}} \right] , \nonumber
\\
\mathrm{and} \hspace{10pt} \lambda (s_{\gamma\gamma}, m_{B}^2, m_{K^*}^2 )
& = & 4 \left( (p_B.p_{K^*})^2 - m_{K^*}^2m_B^2 \right) .
\end{eqnarray}
We have similar expressions for the $\eta$ and $\eta'$ terms,
replacing the parameters $B_{\eta_c}$,
$m_{\eta_c}$ and $\Gamma_{total}^{\eta_c}$ by
their $\eta$ and $\eta'$ counterparts.

\par The irreducible squared matrix element is then;
\begin{eqnarray}
\sum_{spins} \left| M_{irr} \right|^2 & = & \left[ \frac{16
\sqrt{2} \alpha G_F}{9 \pi} V_{tb} V^*_{ts} \right]^2 \frac{q^4}{2
m_{K^*}^2} \left( (p_B.p_{K^*})^2 - m_{K^*}^2m_B^2 \right)
\\
&& \hspace{20pt} \times \left[
\begin{array}{c}
\left| K_{A2} \right|^2 + \left| K'_{A3} \right|^2 + \left|
K'_{A4} \right|^2 + \left| K_{B} \right|^2 + \left| K_{C}
\right|^2 \\
+ 2 \mathrm{Re}(K_{A5}) + 2 \mathrm{Re}(K'_{A3} K^{* \prime}
_{A4}) \\
- 2 \mathrm{Re}(K_{A2} K^{* \prime}_{A3}) - 2
\mathrm{Re}(K_{A2} K^{* \prime}_{A4})
\end{array}
\right] \nonumber
\end{eqnarray}
where
\begin{eqnarray}
K'_{A3} = K_{A3} (m_{B}^2 - m_{K^*}^2) & ; & K'_{A4} = K_{A4} q^2
\nonumber \\
K_{A5}^* & = & K^*_C \left[ \sum_q A_q J(m_q^2) \right] i m_{K^*}
A_0 (q^2)
\end{eqnarray}
The total decay rate is then given by
\begin{equation}
\frac{d\Gamma}{d(\cos\theta)d\sqrt{s_{\gamma\gamma}}} = \sqrt{s_{\gamma\gamma}}
\left(\frac{1}{512 m_B \pi^3}\right) \left[ \left(1 -
\frac{s_{\gamma\gamma}}{m_B^2} + \frac{m_{K^*}^2}{m_B^2} \right)^2 -
\frac{4 m_{K^*}^2}{m_B^2} \right]^{1/2} \sum_{spins} \left| {\cal
M} \right|^2 ,
\end{equation}
where $\sqrt{s_{\gamma\gamma}}$ is the C.M. energy of the two photons while
$\theta$ is the angle which the decaying B-meson makes with the
two photons in the $\gamma \gamma$ C.M. frame.  

\par Our results are presented in figures 1 and 2 (both plotted with a
logarithmic scale on the y-axis), where figure 1
shows the differential branching ratio given as a function of the invariant
mass of the two photons, for the case of the neutral B meson
decay without the interference terms between the resonances
and the irreducible background included. 
The inclusion of the interference terms will in principle give 
rise to an interference
pattern near the base of the resonance peaks. 
Since the peaks are narrow and moreover the interference contributions being
small (except the $\eta_c$-Irreducible term), we do not include them in the
plots. It is worth mentioning
that imposition of suitable cuts in the spectrum to eliminate the
resonance contributions will eliminate any such interference patterns
also. The numerical estimate for the branching ratio arising due to
all possible contributions is summarized in Table 1. Quite evidently,
the largest contribution comes from the $\eta$ resonance mode. It should be
stressed again that using appropriate cuts in the spectrum, the
resonances can be completely eliminated and what is left is the
background irreducible contribution. In estimating the numerical
values for the interference terms, we have assumed that the relative
signs between the terms are such that the $\eta$-Irreducible and 
$\eta^{\prime}$-Irreducible interfering contributions add on to the
other pieces. However, because of the smallness of these values, it
really makes no significant difference.  
\begin{table}[ht]
\begin{center}
\begin{tabular}{|c|c|}
\hline \\
  Contribution & Branching ratio $\times 10^{-7}$ \\
\hline \\
  Resonance &  \\
  $\eta_c$ & 4.7 \\
  $\eta$ & 56.9 \\
  $\eta^{\prime}$ & 3.7 \\\\
  Irreducible & $2.3\times 10^{-2}$ \\\\
  Interference &\\
  $\eta_c$-Irreducible & 2.6\\
  $\eta$-Irreducible & $3\times 10^{-3}$\\
$\eta^{\prime}$-Irreducible & $4.5\times 10^{-3}$\\ \\
BR & 57.5\\ \hline
\end{tabular}
\end{center}
\caption{Contribution to the $B^0 \to K^{* 0} \gamma \gamma$
branching ratio. For the interference terms we quote the absolute values.}
\end{table}

\par In figure 2, we compare the irreducible contribution with the total
contribution to the branching fraction. Clearly, the resonances dominate the
results. At this point, it may be worth mentioning that a quick look at the
individual values tabulated in Table1 reveal the following. Since the
resonances are narrow, one may try to estimate the contributions directly by
multiplying the individual branching ratios ie we expect in the narrow width
approximation that $BR(B-->K^*\gamma\gamma) \sim BR(B-->K^*X)
BR(X-->\gamma\gamma)$ where $X$ denotes any of the resonances. The numbers
quoted clearly show that they are in accord with the expectations.
However, if we had used the form as in Eq(17), we might have over- or
under-estimated the resonance contributions (except probably for the $\eta$
mode) because it is not very clear if such a simple parametrization is the
correct one. We however avoid any such possible conflict by drawing heavily on
the experimental values (upper limit for $\eta'$) of the various sub-process 
branching fractions.  

\par The central values of the parameters used in our calculation are shown in 
the Appendix. We make an attempt to estimate the errors creeping into
the numerical calculations due to errors in various input parameters.
The theoretical uncertainties arising out of uncertainties in the 
parameters are overwhelmingly in the input values of the form factors and 
the meson decay constants . The Wilson coefficients have been taken to be 
their NNL values and there are no significant theoretical uncertainties in 
them. In evaluating the uncertainties of our results, it is appropriate to 
evaluate them separately for the background irreducible contribution and 
the resonance contributions, since the latter can easily be experimentally 
separated from the former by suitable cuts in the spectrum. For the
model dependent parametrization of Eq(17), the 
theoretical uncertainty in the resonance contribution due to $\eta_c$ 
arises mostly because of uncertainties in 
the CKM parameters,
$f_{\eta_c}$, $B_{\eta_c}$ and $F_0(m_{\eta_c^2})$; the Wilson
coefficients values used  are  the NNL level values and no comparable 
uncertainties exist therein. The CKM parameters relevant to us have an
uncertainty of about 10$\%$ \cite{pdg}.
The form factors used are the same as in \cite{Cheng-1}
where the actual dependence of the form factors as a function of momentum 
transfer squared are given and hence no errors arise due to 
parametrization of form factors as a function of $q^2$. Although this 
reference does not quote any estimate of the uncertainties in the
numbers, a typical 
uncertainty in this type of calculation based on quark model is given in 
\cite{beneke} 
and is typically of the order of 15$\%$ ,arising to a great extent due to
uncertainty in the strange quark mass. The rate of the $\eta_c$
decay into two photons is uncertain by about 40$\%$ \cite{pdg}.
A typical estimate of the uncertainty
in the value of $f_{\eta_c}$( arising mostly again out of uncertainty in the 
current mass of the s-quark ) has been estimated at about 15$\%$ in 
\cite{deshpande}.
Combining all this , we would expect that an estimate of the
contribution of the $\eta_c$ resonance based on such a parametrization
to be uncertain by about 50$\%$.
A similar estimate for the other two resonaces give a somewhat lower 
value mostly because their decay rates into two photons is better known,
to an accuracy of about 10 $\%$ for $\eta^{\prime}$ and about 5 $\%$ for the
$\eta$. The uncertainties in the decay constansts of the $eta^{\prime}$
and the $\eta$ have been estimated to be about 10$\%$ \cite{chen}
and we estimate the overall uncertainty in our
calculation for the $\eta^{\prime}$ and $\eta$ to be about 40$\%$ and
30$\%$ respectively. However, since we have relied on experimental 
values of rates and
branching fractions, the above mentioned uncertainties are significantly
reduced and the only source of uncertainty in our estimation is the
uncertainty present in sub-process rates. 

\par Turning to the irreducible contribution, the uncertainty arises mostly
because of the CKM factors and the form factors. These combine to give
an overall uncertainty of about 20$\%$ for the irreducible part of the
amplitude. As stressed before, once the suitable cuts are imposed in
the two photon spectrum, it is possible to extract the irreducible
contribution and here the errors are relatively smaller and are
expected to even go down further with more accurate determination of
the  CKM parameters and the form factors. 
 
\par At the levels reached by the current B-factories, the branching 
ratios obtained are too low to be observed. One certainly hopes that in the 
near future experiments, with better luminosities possible,  
the numbers obtained will be very useful for confronting theoretical 
models with experimental data. As discussed in the text, this decay with 
two photons depends on the parts of the effective Hamiltonian, which
the decays 
with a single photon are not sensitive to and thus provides a more complete 
test of the underlying theory.

\vskip 1.5cm
{\it{\bf Acknowledgements}}~~
SRC would like to acknowledge the Department of Science and Technology,  
Government of India for a research grant.\\
We would like to thank Prof. G.Hiller for a communication pointing out the
discrepancies between their results and our earlier estimates for the
irreducible contribution.
\section{Appendix}
We list the central values of the various parameters entering our
numerical estimates:\\
\[
G_F = 1.16  x 10^{-5} Gev^{-2} \hskip 1cm \alpha(m_B) = 1/130 
\hskip 1cm m_b= 4.8 Gev\]
\[ m_t= 175 Gev \hskip 1cm m_c = 1.5 Gev \hskip 1cm m_u = 0 = m_d \]
\[ m_{K^{*+}}=0.89 Gev \hskip 1cm m_{K^{*0}}=0.896 Gev \]
\[ F_0^{BK^*}=0.3 \hskip 1cm M_{pole}=6.65 Gev \]
\[ \Gamma_{tot}(\eta_c) =1.3 \times 10^{-2}\hskip 1cm
m_{\eta_c} = 3.0 Gev \]
\[ B_{\eta_c} = 2.74 \times 10^{-3} Gev^{-1}\hskip 1cm
f_{\eta_c} = 0.35 Gev \]
\[ \Gamma_{tot}(\eta^{\prime}) = 0.203 \times 10^{-3} Gev \hskip 1cm
m_{\eta^{\prime}} = 0.96 Gev \]
\[ B_{\eta^{\prime}} = 14.0 \times 10^{-3}  Gev ^{-1}\hskip 1cm
f_{\eta^{\prime}}= - 6.3 Mev \]
\[ B(\eta^{\prime} \rightarrow 2 \gamma ) =2.11 \% \hskip 1cm
B(\eta_c \rightarrow 2 \gamma ) = 3 \times 10^{-4}\]

\begin{figure}[ht]
\includegraphics[angle=270,width=15cm]{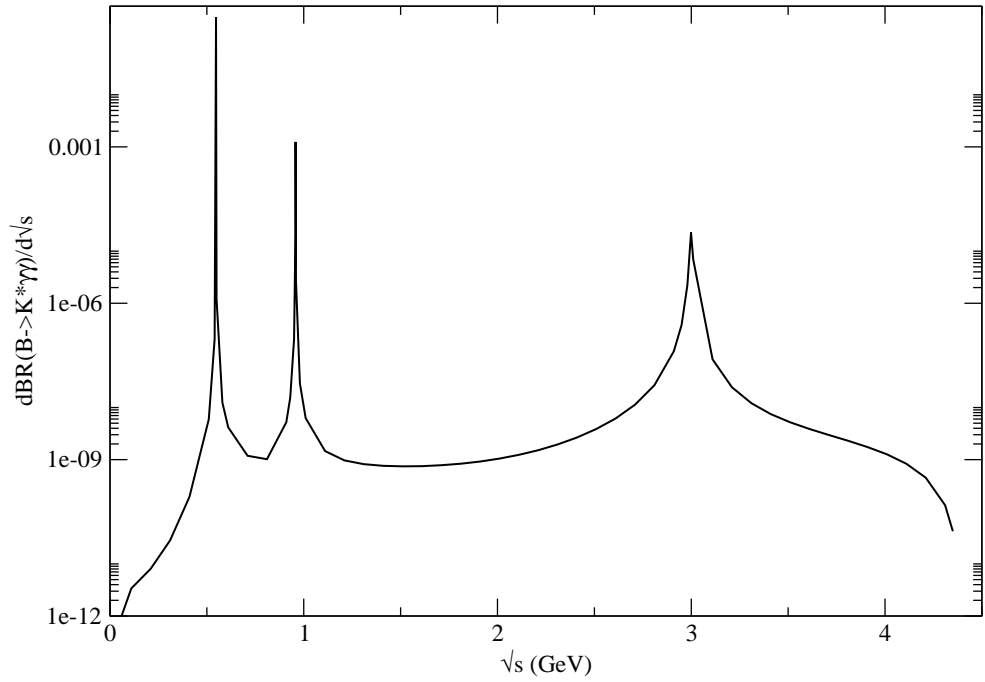}
\caption{The differential branching ratio of $B \to K^* \gamma \gamma$ plotted
as a function of the CM energy of the diphoton rest frame without
interference terms taken into consideration.  Here
we have plotted with a log scale on the y-axis.}\label{fig1}
\end{figure}

\begin{figure}
\includegraphics[angle=270,width=15cm]{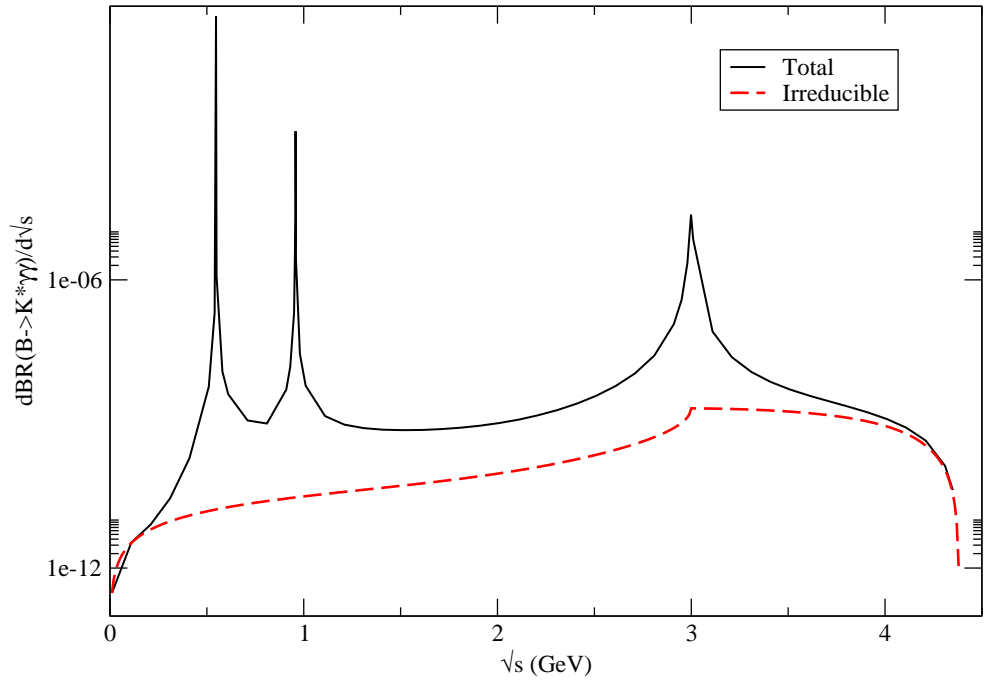}
\caption{The irreducible and the total contribution to the 
differential branching ratio of $B \to K^* \gamma \gamma$ plotted
as a function of the CM energy of the diphoton rest frame.}\label{fig1}
\end{figure}
\end{document}